\documentclass[review]{elsarticle}

\usepackage{lineno,hyperref}
\usepackage{multirow, color}
\usepackage{makecell,amsmath}
\modulolinenumbers[5]
\newcommand\blackf[1]{#1}

\journal{Journal of Pattern Recognition}

\begin{document}

\begin{frontmatter}

\title{SceneFake: An Initial Dataset and Benchmarks for Scene Fake Audio Detection}

\author[1]{Jiangyan Yi\corref{mycorrespondingauthor}\fnref{myfootnote}}
\cortext[mycorrespondingauthor]{\textcolor[rgb]{0,0,0}{Corresponding author, Jiangyan Yi is the corresponding author.}}
\ead{jiangyan.yi@nlpr.ia.ac.cn}

\author[1,2]{Chenglong Wang\fnref{myfootnote}}

\author[3,4]{Jianhua Tao}

\author[1]{Chu Yuan Zhang}

\author[5]{Cunhang Fan}

\author[1]{Zhengkun Tian}

\author[1]{Haoxin Ma}

\author[1]{Ruibo Fu}

\address[1]{\textcolor[rgb]{0,0,0}{State Key Laboratory of Multimodal Artificial Intelligence Systems, Institute of Automation, Chinese Academy of Sciences}}
\address[2]{University of Science and Technology of China}
\address[3]{Department of Automation, Tsinghua University}
\address[4]{Beijing National Research Center for Information Science and Technology,\\Tsinghua University}
\address[5]{Anhui University}
\fntext[myfootnote]{Jiangyan Yi and Chenglong Wang contribute equally to this work.}

\begin{abstract}
  Many datasets have been designed to further the development of fake audio detection.
  However, fake utterances in previous datasets are mostly generated by altering timbre, prosody, linguistic content or channel noise of original \blackf{audio}. \blackf{These datasets leave out a scenario, in which the acoustic scene of an original audio is manipulated with a forged one.} It will pose a major threat to our society if some people misuse the manipulated audio with malicious purpose. Therefore, this motivates us to fill in the gap. This paper \blackf{proposes} such a dataset for scene fake audio detection \blackf{named SceneFake, where a manipulated audio is generated by} only tampering \blackf{with} the acoustic scene of an \blackf{real} utterance by using speech enhancement technologies. \blackf{Some scene fake audio detection benchmark results on the SceneFake dataset are reported in this paper.} \blackf{In addition,} an analysis of fake attacks with different speech enhancement technologies and signal-to-noise ratios are presented \blackf{in this paper.} \textcolor[rgb]{0,0,0}{The results indicate that scene fake utterances cannot be reliably detected by baseline models trained on the ASVspoof 2019 dataset. Although these models perform well on the SceneFake training set and seen testing set, their performance is poor on the unseen test set.} \blackf{The dataset~\footnote{https://zenodo.org/record/7663324\#.Y\_XKMuPYuUk} and benchmark source codes~\footnote{https://github.com/ADDchallenge/SceneFake} are publicly available.}
\end{abstract}


\end{frontmatter}

\section{Introduction}

\begin{figure}[htb]
\hfill
\begin{minipage}[b]{1.0\linewidth}
  \centering
  \centerline{\includegraphics[width=9.6cm,height=5.7cm]{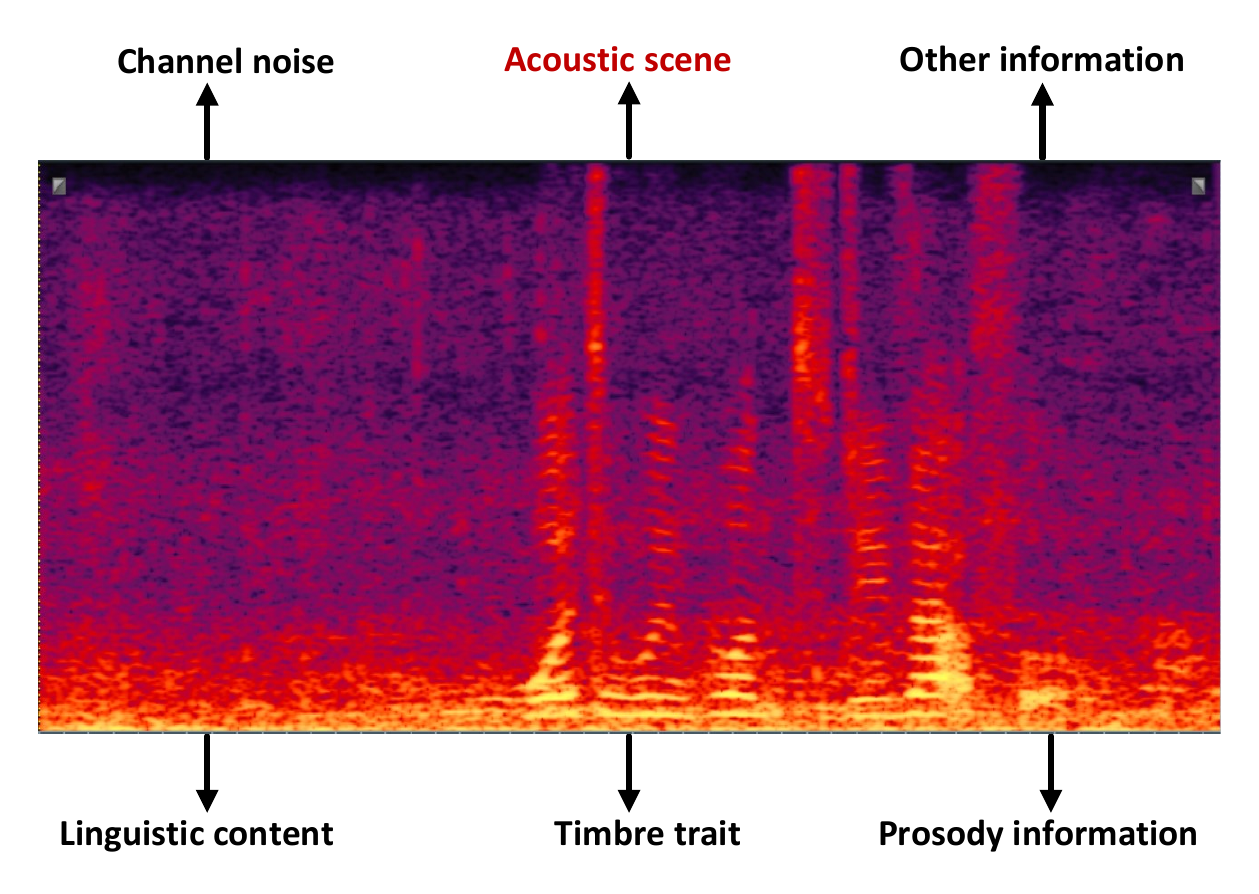}}
\end{minipage}
\caption{Spectrogram of an example utterance ``\textit{What do we want to do that for?}''. It involves rich information, such as timbre trait, prosody feature, linguistic content, channel noise, acoustic scene and other information. The acoustic scene of the utterance is \textit{Airport}.}
\label{fig:speech}
\end{figure}

Speech signals contain rich information in real life scenarios. A spectrogram of an example utterance is shown in Figure~\ref{fig:speech}. It involves not only timbre trait, prosody feature, linguistic content and channel noise but also acoustic scene and \textcolor[rgb]{0,0,0}{other contextual information}. Acoustic scene is a kind of acoustic environments~\cite{Zhao2013Aud} \blackf{that describes the location of happened events recorded} in the audio, such as bus, park, airport, metro station and so on~\cite{Ma2006Acou}. \blackf{As an example, the acoustic scene of the utterance in Figure~\ref{fig:speech} is \textit{Airport}.} If the scene of an original audio is manipulated with another scene, \blackf{the} authenticity and integrity verification of the audio will be unreliable and even the semantic meaning of the original audio \blackf{may be changed given the shift in context.}
The goal of speech enhancement is to remove noise signals and estimate a target clean speech from a noisy audio. During the past few years\blackf{,} speech enhancement technologies \blackf{have} made significant progress with the development of deep learning~\cite{pandey2019n}.
\blackf{Current speech enhancement models are able to effectively suppress the acoustic scene of an audio sample.}
The intelligibility and quality of the enhanced audio are very \blackf{close} to \textcolor[rgb]{0,0,0}{ the clean one~\cite{Pegg2024RTFSNetRT}, demonstrating the ease and effectiveness of manipulating} the scene of an utterance with another scene using speech enhancement technologies.

\blackf{This is a cause for concern, therefore, if someone misuses} scene manipulated \blackf{audio} with the intent to cause harm.
For one thing, \blackf{it would \textcolor[rgb]{0,0,0}{mislead the public} if scene-manipulated audio was disseminated widely on social media} for unethical purposes. \textcolor[rgb]{0,0,0}{For another, the scene manipulated audio increases the risk of attacks for many applications~\cite{Stowell2015Detection} including intelligent wearable devices, context-aware services, robotics navigation systems, \blackf{which rely on acoustic scene classification/recognition to understand the situation of their users.} \blackf{Scene fake utterances also bring potential vulnerability risk \textcolor[rgb]{0,0,0}{to audio-visual scene analysis systems~\cite{Owens2018AV} and objects in the environment~\cite{Gan2019Self} using both visual and acoustic cues,} as well as recognizing environmental sound via audio-visual learning~\cite{Zhao2018TheSO}.} \blackf{Furthermore,} the acoustic scene manipulation technology \blackf{makes} real-time crime localization systems unreliable~\cite{Malik2013Acoustic}. For example, a victim being chased by an offender calls a police emergency call center. The actual location (i.e. car, street, living room etc.) of harassed persons may not be \blackf{correctly identified} if the localization system is attacked by the \blackf{acoustic} scene manipulation technology. In addition, it will bring challenges to audio forensics. Audio forensics \blackf{seeks} to evaluate and \blackf{analyze} audio recordings, which is commonly
utilized for integrity verification and authenticity of the evidence in a court of law~\cite{Zakariah2017D}. One scenario of audio forensics is to identify and rebuild crime or accident scenes. It may pose serious risks if real scenes of the recordings are \textcolor[rgb]{0,0,0}{replaced} with a fake one.
\textcolor[rgb]{0,0,0}{Therefore,} scene manipulation audio detection is of great significance. It is also nontrivial to design a scene fake audio dataset to help the development of this research.}

\begin{figure}[tb]
\hfill
\begin{minipage}[b]{1.0\linewidth}
  \centering
  \centerline{\includegraphics[width=9.7cm,height=5.5cm]{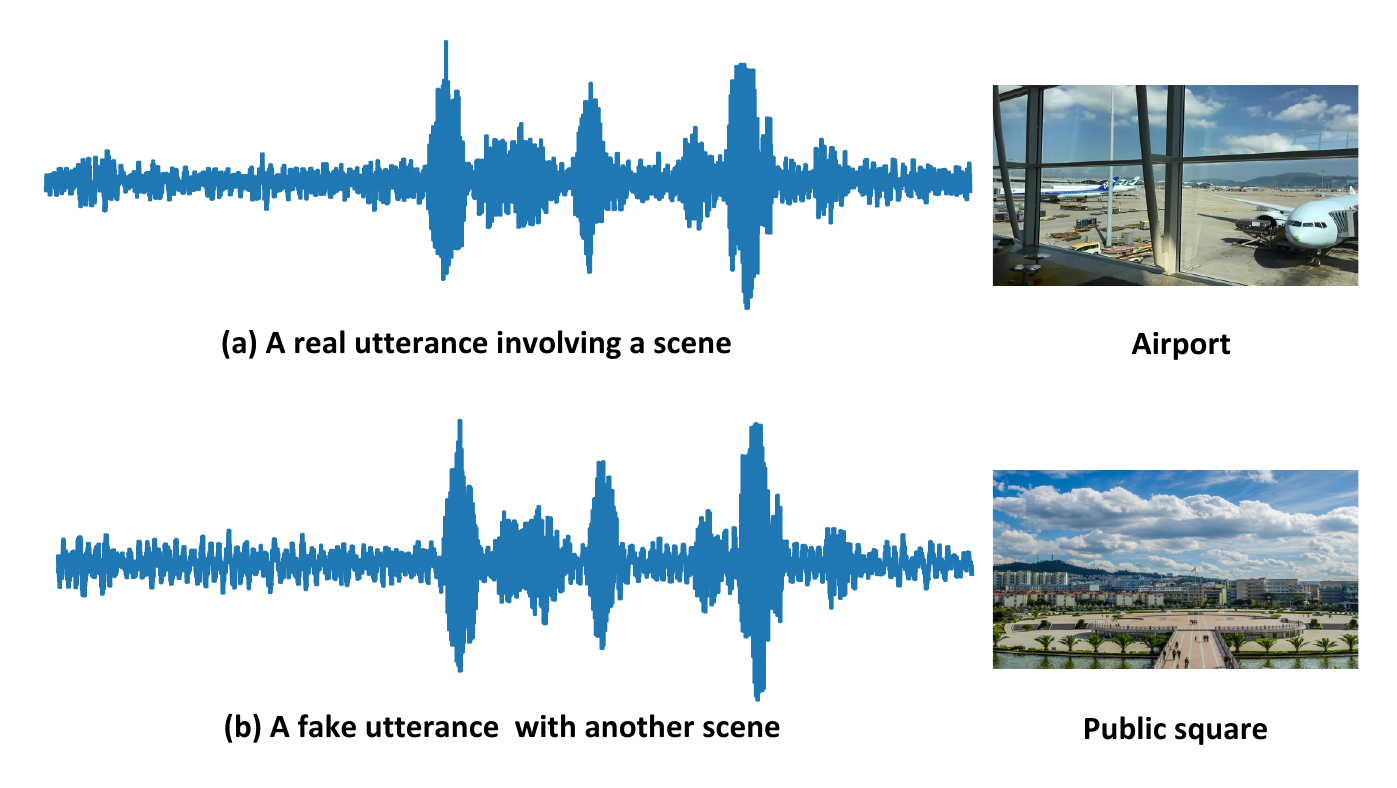}}
\end{minipage}
\caption{Waveforms of example utterances. (a) illustrates a real utterance involving a scene, such as ``\textit{Airport}''. (b) shows a fake utterance: the scene of the real utterance is manipulated with another scene, such as ``\textit{Public square}''.}
\label{fig:example}
\end{figure}

\blackf{A growing number of datasets have been built to promote research on detecting fake utterances.}
Most of previous datasets are focused on detecting spoofed utterances for automatic speaker verification (ASV) systems. There are about four types of spoofing attacks~\cite{Wu2015Spoofing}: impersonation, replay, speech synthesis and voice conversion.
\blackf{A few} datasets are designed for deepfake audio detection including in the ASVspoof 2021~\cite{2021ASVspoof}\textcolor[rgb]{0,0,0}{, the first Audio Deep synthesis Detection (ADD 2022)~\cite{Yi2022ADD}challenge~\footnote{http://addchallenge.cn/add2022} and the second Audio Deepfake Detection (ADD 2023) ~\cite{Yi2023ADD} challenge~\footnote{http://addchallenge.cn/add2023}} \blackf{considering some ignored challenging fake scenarios in real life.} \blackf{Deepfake} utterances are roughly classified into three types: speech synthesis, voice conversion and speech manipulation.
Impersonation~\cite{Wu2015Spoofing} denotes human mimicking \blackf{where} an imitator mimics the voice timbre and prosody of a target speaker.
Replay attack~\cite{Kinnunen2017ASVspoof} \blackf{refers to} a form of replaying pre-recorded bona fide utterances of a target speaker to an ASV system. An example is the recording replayed using a smart device.
Speech synthesis~\cite{Wu2015Spoofing} is a technique for generating intelligible and natural speech for any arbitrary text using machine learning based models.
Voice conversion~\cite{Wu2015Spoofing} \blackf{aims to change the timbre and prosody of the utterance so that it matches the style of another speaker via computer-aided technologies.}
Speech manipulation~\cite{Yi2021Half} \blackf{involves generating} partially fake utterances by manipulating the original genuine utterances with bona fide or synthesized audio segments.

The \blackf{aforementioned datasets} are crucial \blackf{in} accelerating research on detecting fake utterances. The datasets of \textcolor[rgb]{0,0,0}{ASVspoof 2021~\cite{2021ASVspoof} and ADD2022~\cite{Yi2022ADD} and ADD 2023~\cite{Yi2023ADD} challenges} have especially played a significant role in promoting the development of this research.
\textcolor[rgb]{0,0,0}{However, the fake utterances in previous datasets are mainly generated by changing timbre, prosody, linguistic content or channel noise of the original utterance. They have not covered a fake situation which involves manipulating the acoustic scene of the original audio with the another one. An example of scene manipulated audio is shown in Figure~\ref{fig:example} (b). The scene ``\textit{Airport}'' is \blackf{replaced} with another scene~``\textit{Public square}''. Such change and replacement may affect the semantics of the utterance, which \textcolor[rgb]{0,0,0}{rely heavily} on context. Therefore, this paper is motivated to fill in the gap.}

We report our progress in developing such a scene manipulated corpus involving changing the scene of an audio with another scene using speech enhancement technologies. The dataset is named \blackf{Scene Fake Audio Detection} (SceneFake) dataset. We describe a comparison of several scene manipulation detection methods to discriminate between the genuine and fake speech. A preliminary set of benchmark results for detecting fake utterances is presented in this paper. Furthermore, we conduct an analysis of fake attacks with various speech enhancement technologies on our designed SceneFake dataset.

The main contributions of this paper are as follows.
To the best of our knowledge, this is the first attempt to \blackf{shed light on} such an audio fake attack using speech enhancement technologies and design a scene manipulated audio dataset for fake audio detection. The SceneFake dataset provides the speech enhancement technology information of the fake utterances. In addition, the dataset includes seen and unseen test sets. Researchers can evaluate the performance and generalization of a fake detection model.
\blackf{The SceneFake dataset is publicly available}.

The rest of this paper is organized as follows. Section 2 introduces the related work. Section 3 describes the design policy of the dataset. Evaluation metric is introduced in Section 4. Section 5 presents experiments and baselines. Section 6 discusses the results and future work. This paper is concluded in Section 7.

\section{Related Work}

\subsection{Fake Audio Detection Datasets}
Most of \blackf{the} previous spoofed datasets \blackf{focus} on developing countermeasures for automatic speaker verification systems, which mainly include four kinds of spoofing types~\cite{Wu2015Spoofing}: impersonation, replay, speech synthesis and voice conversion.
In 2004, an impersonation database \blackf{was} developed by Lau et al.~\cite{Lau2004Impdataset}, which is used for investigating the vulnerability of speaker verification. In 2013, a small Finnish impersonation dataset \blackf{was} designed by Hautamaki et al.~\cite{Hautamaki2013Impdataset}.
A few individual spoofing datasets \textcolor[rgb]{0,0,0}{have been designed for speaker verification systems, where the spoofing types only involve a kind of speech synthesis method or a sort of voice conversion approach.} The spoofing types are not diverse. In 2015, a standard spoofing database SAS is designed by Wu et al.~\cite{Wu2015SAS}, which is \blackf{composed} of various speech synthesis and voice conversion methods. The SAS database is used for supporting the first ASVspoof challenge~\cite{Wu2015ASVspoof} \blackf{, organized} for detecting the spoofed speech in 2015. \blackf{Replay is included in the dataset of the ASVspoof 2017 challenge~\cite{Kinnunen2017ASVspoof} due to lowcost and challenging.} The ASVspoof 2019 database~\cite{ASVspoof2019} is comprised of replay, speech synthesis and voice conversion attacks. Previous datasets in ASVspoof challenges aim to detect unforseen attack in microphone channel. Lavrentyeva et al.~\cite{Galina2019Phone} design a PhoneSpoof dataset for speaker verification systems, in which the utterances are collected in telephone channels.

Recently, a few attempts have been made to develop datasets mainly for fake audio detection systems. Reimao et al.~\cite{Reimao2020For} design a dataset for synthetic speech detection. The fake utterances \blackf{are} generated by the open-sourced tools only using the latest speech synthesis technology. Frank et al.~\cite{Frank2021WaveFake} develop a fake dataset named WaveFake, which contains fake utterances generated by the latest speech synthesis models. ASVspoof 2021~\cite{2021ASVspoof} \blackf{includes audio deepfake attacks, replay, speech synthesis and voice conversion spoofing methods.} However, these datasets have not covered many real-life challenging situations. The ADD 2022 challenge was motivated to fill the gap~\cite{Yi2022ADD}. \blackf{The ADD 2022 consists of various datasets including fully fake utterances containing various noises, partially fake utterances, and adversarial examples.} \textcolor[rgb]{0,0,0}{However, the ADD 2023~\cite{Yi2023ADD} focuses on surpassing the constraints of binary discrimination, and actually localizing the manipulated intervals in a partially fake speech as well as pinpointing the source responsible for generating any fake audio. Most recently, Zang et al~\cite{zang2024singfake} present a dataset named SingFake containing deepfake song clips.}

The above-mentioned datasets have played a key role in accelerating the development of anti-spoofing and audio deepfake detection. However, the fake utterances in these datasets mainly involve changing timbre, prosody, linguistic content or channel noise of the original audio. They do not consider the \blackf{manipulation of the acoustic scene of the original audio with a forged one.}

\begin{table}[!t]
\caption{The description of acoustic scenes in the DCASE 2022 Challenge.}
\centering
\label{tab:scene}
\begin{tabular}{|c||c|}
\hline
Scene & Description \\
\hline
Airport & Airport\\
Bus & Travelling by a bus\\
Park & Urban park\\
Public  & Public square\\
Shopping  & Indoor shopping mall\\
Station & Metro station\\
Metro & Travelling by an underground metro\\
Pedestrian & Pedestrian street \\
Street & Street with medium level of traffic\\
Tram & Travelling by a tram\\
\hline
\end{tabular}
\end{table}

\subsection{Acoustic Scene Classification Corpus}
Acoustic scene classification has a wide range of applications\blackf{,} ranging from audio recording integrity authentication to real-time crime identification~\cite{Malik2013Aco}. It attempts to recognize acoustic scene labels of audio signals, such as an airport or a park environment.
A diverse set of \blackf{corpora have been} designed to identify acoustic scenes.
There are a series of available datasets for acoustic scene classification, which are provided by the detection and classification of acoustic scenes and events (DCASE) challenges~\footnote{https://dcase.community}. \blackf{The acoustic scene classification task of DCASE 2020 is a classification of data from multiple devices (real and simulated) targeting generalization properties of systems across a number of different devices~\cite{Heittola2020}. The acoustic scene classification dataset of DCASE 2022 contains recordings in 10 kinds of acoustic scenes collected from 12 European cities with 4 different devices~\cite{heittola2022TAU}.} \textcolor[rgb]{0,0,0}{The task of DCASE 2023~\footnote{https://dcase.community/challenge2023/index} is a repeat task from DCASE 2022, which update the evaluation metrics by adding calculation of the energy consumption.}

These datasets are \blackf{vital} for acoustic scene classification tasks.
The output labels of the acoustic scene classification system will become \blackf{unreliable}
if the scene of the original audio is tampered \blackf{with} speech enhancement technologies.
Different from these datasets, this paper aims to design \blackf{an acoustic} scene manipulated dataset named SceneFake \blackf{to further promote} research on fake audio detection.

\section{Dataset Design}

Our scene fake audio detection (SceneFake) dataset is developed based on the logical access (LA) dataset of ASVspoof 2019 and the acoustic scene dataset from DCASE 2022 challenge.
The LA dataset~\footnote{https://datashare.ed.ac.uk/handle/10283/3336} consists of genuine and spoofed audio segments involving synthetic utterances and converted voices.
It consists of three sets: training, development and test set. The acoustic scene dataset~\footnote{https://zenodo.org/record/6337421} contains 64 hours of 10-seconds audio segments from 10 acoustic scenes.
The description of 10 acoustic scenes in the DCASE 2022 Challenge is reported in the Table~\ref{tab:scene}. The statistics of the LA dataset of ASVspoof 2019 are listed in Table~\ref{tab:la}, \blackf{where \#Speakers, \#Genuine, \#Spoofed, and \#Total denote the number of speakers, genuine utterances, spoofed utterances, and all utterances in the three sets of LA dataset.}

\subsection{Design Policy}
\begin{figure*}[htb]
\hfill
\begin{minipage}[b]{1.0\linewidth}
  \centering
  \centerline{\includegraphics[width=12.5cm,height=2.6cm]{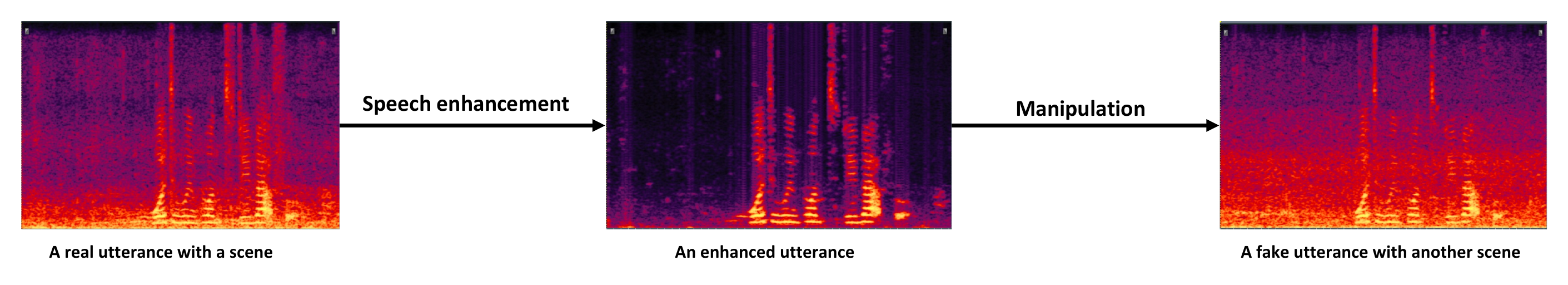}}
\end{minipage}
\caption{An example of acoustic scene manipulation for a fake utterance. The manipulation procedure consists of two steps: 1. Enhancing the real speech involving a scene, such as ``\textit{Airport}''. 2. Adding another scene to the enhanced speech, such as ``\textit{Street}''. The signal noise ratio (SNR) of the real utterance is denoted by \blackf{SNR$_{real}$}. The SNR of the fake utterance \blackf{ is referred to as} \blackf{SNR$_{fake}$}. The \blackf{SNR$_{real}$} and \blackf{SNR$_{fake}$} are both 5dB in the example.}
\label{fig:manipulation}
\end{figure*}

The SceneFake dataset is designed to evaluate and analyze the methods of detecting scene manipulated utterances.
The dataset consists of genuine and fake utterances involving \blackf{different} scenes.
\textcolor[rgb]{0,0,0}{The essence of the SceneFake dataset lies in the acoustically scene-manipulated audio, which is identified as a fake utterance.}
An example of acoustic scene manipulation for \blackf{a fake audio} is illustrated in Figure~\ref{fig:manipulation}.
The generation procedure of manipulation \textcolor[rgb]{0,0,0}{comprises} two steps:
\begin{enumerate}
\item{Enhancing the real speech involving a scene}
\item{Adding another scene to the enhanced speech}
\end{enumerate}
The SNR of the real utterance and the fake utterance are denoted by \blackf{SNR$_{real}$} and  \blackf{SNR$_{fake}$}, respectively.
In order to simplify the problem, The \blackf{SNR$_{real}$} is identical to the corresponding \blackf{SNR$_{fake}$} for the manipulation in this work. For instance, the \blackf{SNR$_{real}$} and \blackf{SNR$_{fake}$} are both 5dB in Figure~\ref{fig:manipulation}.

\subsection{Real Utterance Collection}

In the field of speech enhancement, researchers~\cite{Tan2017Speech} usually simulate real noisy utterances by mixing clean utterances with noise signals at various signal-to-noise ratios (SNRs) to overcome the difficulty in obtaining clean utterances corresponding to noisy ones. This inspires us to generate a noisy speech by mixing a clean utterance with \blackf{an acoustic} scene. A simulated noisy speech is viewed as a real utterance involving a scene in the SceneFake dataset.

We add a variety of scenes to clean utterances to simulate genuine utterances.
The clean \blackf{utterances} come from bona fide utterances of the LA dataset in \blackf{ASVspoof 2019~\cite{ASVspoof2019}. The bona fide utterances are from a multi-speaker English speech database recorded in a hemi-anechoic chamber under clean conditions, which are downsampled to 16 kHz at 16 bits-per-sample.} The acoustic scenes come from the dataset in DCASE 2022 challenge as shown in Table~\ref{tab:scene}.
The real utterances of our training, development and test sets are generated based upon the bona fide ones of training, development and test sets from the LA dataset, respectively. They are generated by randomly adding acoustic scenes to clean utterances at 6 different \blackf{SNR$_{fake}$} each -5dB, 0dB, 5dB, 10dB, 15dB and 20dB.

\subsection{Scene Manipulation for Fake Audio}

Scene manipulation consists of two steps as shown in Figure~\ref{fig:manipulation}. The first step is to remove the scene of the simulated real speech using speech enhancement technologies. In the last step the enhanced speech is added with another scene. The scene manipulation for a fake speech is represented as:
\begin{align}\label{eq-fake}
\widehat{x}_{enhanced}(t)&=\text{SE}(x_{real}(t)) \\
y_{fake}(t)&=\widehat{x}_{enhanced}(t) + n_{scene}(t)
\end{align}
where $\text{SE}(\cdot)$ denotes the speech enhancement function, which aims to estimate the clean target speech from the real speech $x_{real}(t)$, $\widehat{x}_{enhanced}(t)$ is the enhanced speech, $n_{scene}(t)$ denotes target acoustic scene, $y_{fake}(t)$ \blackf{ is referred to as} the fake audio whose scene is manipulated with another one.

\begin{figure*}[htb]
\hfill
\begin{minipage}[b]{1.0\linewidth}
  \centering
  \centerline{\includegraphics[width=12.5cm,height=4.8cm]{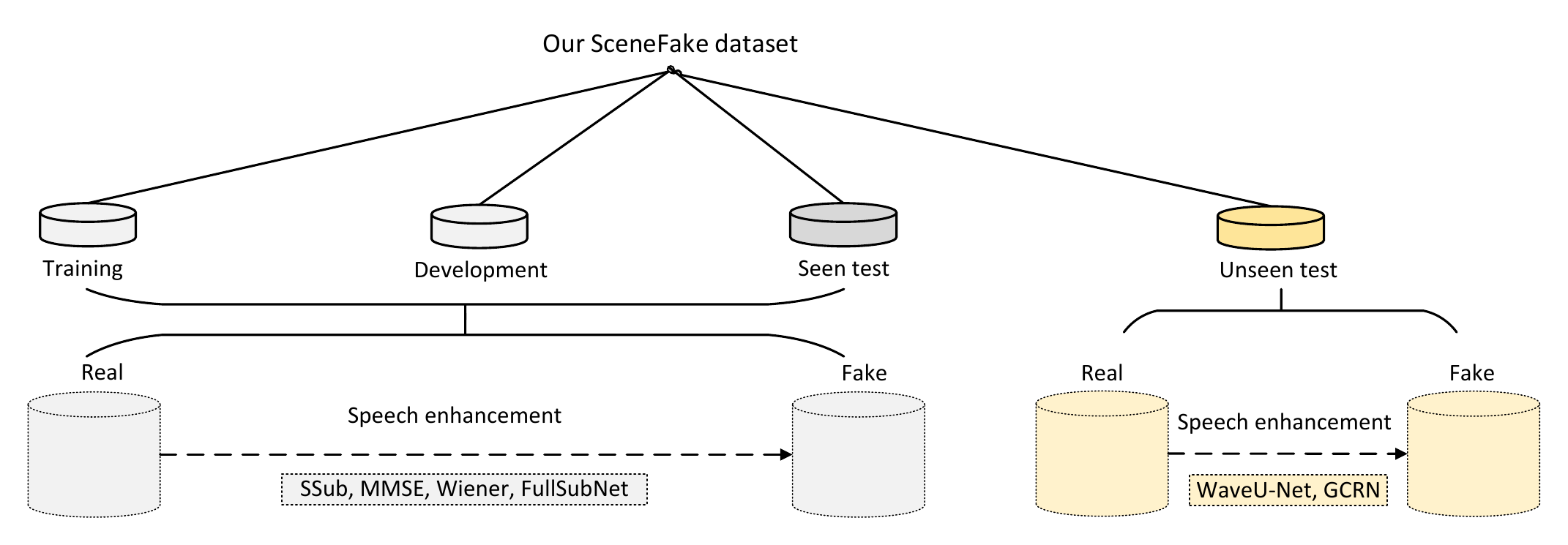}}
\end{minipage}
\caption{Data structure of the SceneFake dataset. It consists of five sets: training, development, seen test, unseen test~1 and unseen test~2 sets.}
\label{fig:dataset}
\end{figure*}

We employ several \blackf{recent} state-of-the-art neural network based speech enhancement methods and some traditional open-source speech enhancement models to remove the scene of the real speech.
A commonly used open source speech enhancement model is utilized to enhance the speech. \blackf{The model, named FullSubNet~\cite{hao2021full}, is composed of a full-band and a sub-band fusion model}, which outperforms the top-ranked methods in the deep noise suppression (DNS) challenge~\cite{reddy2020is}. A time-domain based speech enhancement method, named WaveU-Net~\cite{Danieal2018Unet}, is employed to enhance the real audio. The WaveU-Net model is migrated and implemented based on an end-to-end source separation in the time-domain, which \blackf{allows for} modelling phase information and avoids fixed spectral transformations. A gated convolutional recurrent network (GCRN) is employed to remove the scene~\cite{Tan2019GCRN}. The GCRN model obtains better performance than an existing convolutional neural network in terms of both objective speech intelligibility and quality. Furthermore, we compare the performance of neural network based models with other traditional open-source speech enhancement methods~\cite{Loizou2007Speech}, such as spectral subtraction (SSub), minimum mean square error (MMSE), and Wiener filtering (Wiener). The above-mentioned speech enhance models are \blackf{publicly available, ensuring the reproducibility (of our database design).}

The fake utterances are generated by mixing another randomly sampled acoustic scenes with the enhanced utterances \blackf{each mixed with 6 different SNR$_{fake}$} -5dB, 0dB, 5dB, 10dB, 15dB and 20dB. Fake utterances are also generated by using an open-source toolkit Augly.

\begin{table}[!t]
\caption{The statistics of LA dataset of ASVspoof 2019. }
\centering
\label{tab:la}
\begin{tabular}{|c||cccc|}
\hline
Set & \#Speakers & \#Genuine & \#Spoofed & \#Total\\
\hline
Training & 20 &  \textcolor[rgb]{0,0,0}{2,580} & \textcolor[rgb]{0,0,0}{22,800} & \textcolor[rgb]{0,0,0}{25,285}\\
Development & 20 & \textcolor[rgb]{0,0,0}{2,548} & \textcolor[rgb]{0,0,0}{22,296} & \textcolor[rgb]{0,0,0}{24,844}\\
Test & 67 & \textcolor[rgb]{0,0,0}{7,355} & \textcolor[rgb]{0,0,0}{63,882} & \textcolor[rgb]{0,0,0}{71,237}\\
\hline
\end{tabular}
\end{table}

\begin{table*}[!t]
\caption{The statistics of our SceneFake dataset.}
\centering
\label{tab:datasetinfo}
\begin{tabular}{|c||cccccc|}
\hline
Set & \#Speakers & \#SE & \#Scenes & \#Real & \#Fake & \#Total\\
\hline
Training & 20 & 4 & 6 & \textcolor[rgb]{0,0,0}{2,580} & \textcolor[rgb]{0,0,0}{10,320} & \textcolor[rgb]{0,0,0}{12,900}\\
Development & 20& 4 & 6  & \textcolor[rgb]{0,0,0}{2,548} & \textcolor[rgb]{0,0,0}{10,192} & \textcolor[rgb]{0,0,0}{12,740}\\
Seen test & 67 & 4 & 6 & \textcolor[rgb]{0,0,0}{7,355} & \textcolor[rgb]{0,0,0}{29,420} & \textcolor[rgb]{0,0,0}{36,775}\\
Unseen test & 67 & 2 & 4 & \textcolor[rgb]{0,0,0}{7,355} & \textcolor[rgb]{0,0,0}{14,710} & \textcolor[rgb]{0,0,0}{22,065}\\
\hline
\end{tabular}
\end{table*}

\subsection{Datasets Composition}

There are \blackf{four} sets in the SceneFake dataset: training, development, seen test and unseen test. We design \blackf{an unseen test set} to evaluate the generalization of the models. There are no overlaps among the speakers of training, development and seen test set.
\blackf{The speakers of the unseen test set are identical to that of the seen test set. The dataset consists of real and fake utterances with various scenes. Our training, development and seen test sets are populated with utterances with 6 kinds of acoustic scenes: Airport, Bus, Park, Public, Shopping, Station, and the fake utterances therein are manipulated by using four kinds of speech enhancement methods: SSub, MMSE, Wiener, FullSubNet.  Our unseen test test is populated with utterances with four kinds of acoustic scenes: Metro, Pedestrian, Street, Tram, and the fake utterances therein are manipulated by using two kinds of speech enhancement methods: WaveU-Net, GCRN. The acoustic scenes are randomly sampled to mix with the utterances at 6 different SNRs each:} -5dB, 0dB, 5dB, 10dB, 15dB and 20dB.
\textcolor[rgb]{0,0,0}{The data structure and the detailed configurations of acoustic scene manipulation in the SceneFake dataset are illustrated in Figure~\ref{fig:dataset}.}

The statistics of the SceneFake dataset are shown in Table~\ref{tab:datasetinfo}, \blackf{where \#Speakers, \#SE, \#Scenes, \#Real, \#Fake, and \#Total denote the number of speakers, speech enhancement methods, acoustic scene types, real utterances, fake utterances, and all utterances in the SceneFake dataset.}
The statistics of real and fake utterances in our SceneFake dataset at different SNRs are reported in Tables~\ref{tab:datasetinfo-snr1} and ~\ref{tab:datasetinfo-snr2}, \blackf{where \#-5dB, \#0dB, \#5dB, \#10dB, \#15dB and \#20dB denote the number of real or fake utterances at 6 different SNRs each -5dB, 0dB, 5dB, 10dB, 15dB and 20dB.}

\begin{table*}
\caption{The statistics of real utterances in our SceneFake dataset at 6 SNRs.}
\label{tab:datasetinfo-snr1}
\centering
\begin{tabular}{|c||ccccccc|}
\hline
\multirow{2}{*}{\makecell[c]{Set}} &
\multicolumn{7}{c|}{Real}\\
\cline{2-8}
 &  \#-5dB &  \#0dB & \#5dB  & \#10dB & \#15dB  &  \#20dB & \#Total\\
\hline
Training & 430 & 430 & 430 & 430 & 430 & 430 &  \textcolor[rgb]{0,0,0}{2,580} \\
Development & 424 & 424 & 425 & 425 & 425 & 425 &  \textcolor[rgb]{0,0,0}{2,548}\\
Seen test &  \textcolor[rgb]{0,0,0}{1,226} &  \textcolor[rgb]{0,0,0}{1,226} &  \textcolor[rgb]{0,0,0}{1,226} &  \textcolor[rgb]{0,0,0}{1,226} &  \textcolor[rgb]{0,0,0}{1,226} &  \textcolor[rgb]{0,0,0}{1,225} &  \textcolor[rgb]{0,0,0}{7,355}  \\
Unseen test &  \textcolor[rgb]{0,0,0}{1,226} &  \textcolor[rgb]{0,0,0}{1,226} &  \textcolor[rgb]{0,0,0}{1,226} &  \textcolor[rgb]{0,0,0}{1,226} &  \textcolor[rgb]{0,0,0}{1,226 }&  \textcolor[rgb]{0,0,0}{1,225} &  \textcolor[rgb]{0,0,0}{7,355} \\
\hline
\end{tabular}
\end{table*}

\begin{table*}
\caption{The statistics of fake utterances in our SceneFake dataset at 6 SNRs.}
\label{tab:datasetinfo-snr2}
\centering
\begin{tabular}{|c||ccccccc|}
\hline
\multirow{2}{*}{\makecell[c]{Set}} &
\multicolumn{7}{c|}{Fake}\\
\cline{2-8}
 &  \#-5dB &  \#0dB & \#5dB  & \#10dB & \#15dB  &  \#20dB & \#Total\\
\hline
Training &  \textcolor[rgb]{0,0,0}{1,720} &  \textcolor[rgb]{0,0,0}{1,720}&  \textcolor[rgb]{0,0,0}{1,720} &  \textcolor[rgb]{0,0,0}{1,720} &  \textcolor[rgb]{0,0,0}{1,720} & \textcolor[rgb]{0,0,0}{1,720} &  \textcolor[rgb]{0,0,0}{10,320} \\
Development  &  \textcolor[rgb]{0,0,0}{1,696} &  \textcolor[rgb]{0,0,0}{1,696}& \textcolor[rgb]{0,0,0}{1,700} &  \textcolor[rgb]{0,0,0}{1,700} &  \textcolor[rgb]{0,0,0}{1,700} & \textcolor[rgb]{0,0,0}{1,700}& \textcolor[rgb]{0,0,0}{10,192} \\
Seen test  &  \textcolor[rgb]{0,0,0}{2,452} &  \textcolor[rgb]{0,0,0}{2,452} &  \textcolor[rgb]{0,0,0}{2,452} &  \textcolor[rgb]{0,0,0}{2,452} &  \textcolor[rgb]{0,0,0}{2,452} &  \textcolor[rgb]{0,0,0}{2,450} &  \textcolor[rgb]{0,0,0}{14,710} \\
Unseen test  &  \textcolor[rgb]{0,0,0}{2,452} & \textcolor[rgb]{0,0,0}{2,452} &  \textcolor[rgb]{0,0,0}{2,452} & \textcolor[rgb]{0,0,0}{2,452}&  \textcolor[rgb]{0,0,0}{2,452} &  \textcolor[rgb]{0,0,0}{2,450} &  \textcolor[rgb]{0,0,0}{14,710} \\
\hline
\end{tabular}
\end{table*}

\section{Evaluation Metrics}
The goal of audio scene manipulation detection is to develop a method or an algorithm to discriminate between the manipulated audio and the genuine one. \blackf{Motivated by previous evaluation metrics in the ASVspoof challenges and \blackf{the} ADD 2022 challenge, we use textcolor[rgb]{0,0,1}{the} equal error rate (EER)~\cite{Wu2015ASVspoof} and \blackf{the} minimum normalized tandem detection cost function (t-DCF)~\cite{2021ASVspoof} as the evaluation metrics of scene fake audio detection tasks.}

\blackf{The 'threshold-free' EER is used in this work, defined as follows. Let $P_{\mathrm{fa}}(\theta)$ and $P_{\mathrm{miss}}(\theta)$ denote the false alarm and miss rates at threshold $\theta$.}
\begin{align}\label{eer}
P_{\mathrm{fa}}(\theta)&=\frac{\# \{{\textit{manipulated trials with score $\textgreater$ $\theta$}} \}}{\# \{{\textit{total manipulated trials}} \}} \\
P_{\mathrm{miss}}(\theta)&=\frac{\# \{{\textit{genuine trials with score $\textless$ $\theta$}} \}}{\# \{{\textit{total genuine trials}} \}}
\end{align}

\blackf{$P_{\mathrm{fa}}(\theta)$ and  $P_{\mathrm{miss}}(\theta)$ are, respectively, monotonically decreasing and increasing functions of $\theta$.  The EER corresponds to the threshold $\theta_{\mathrm{EER}}$ at which the two detection error rates are equal, i.e. $\mathrm{EER}=P_{\mathrm{fa}}(\theta_{\mathrm{EER}}) =P_{\mathrm{miss}}(\theta_{\mathrm{EER}})$. A lower EER indicates a better performance of the model.}

\blackf{The performance of the scene fake audio detection is also evaluated in terms of t-DCF for the sake of easier tractability, which is defined as follows:
\begin{equation}\label{tdcf}
\mathrm{\mbox{t-DCF}}=\min_{\theta}\left\{\beta P_{\mathrm{miss}}(\theta)+P_{\mathrm{fa}}(\theta)\right\}
\end{equation}
where $\beta$ depends on application parameters (priors, costs), while $P_{\mathrm{miss}}(\theta)$ and $P_{\mathrm{fa}}(\theta)$ are the miss and false alarm rates
at threshold $\theta$. The minimum in Equation~\ref{tdcf} is taken over all thresholds on development or evaluation data. A lower t-DCF also indicates a better performance of the model.}

\section{Initial Benchmarking Experiments}
A series of benchmark experiments are conducted on the SceneFake dataset.

\begin{table}[!t]
\caption{The statistics of the generated noisy LA dataset of ASVspoof 2019.}
\centering
\label{tab:la-noisy}
\begin{tabular}{|c||cccc|}
\hline
Set & \#Speakers & \#Genuine & \#Spoofed & \#Total\\
\hline
Training & 20 & \textcolor[rgb]{0,0,0}{2,580} & \textcolor[rgb]{0,0,0}{22,800} & \textcolor[rgb]{0,0,0}{25,285}\\
Development & 20 & \textcolor[rgb]{0,0,0}{2,548} & \textcolor[rgb]{0,0,0}{22,296} & \textcolor[rgb]{0,0,0}{24,844}\\
Seen test & 67 & \textcolor[rgb]{0,0,0}{7,355} & \textcolor[rgb]{0,0,0}{63,882} & \textcolor[rgb]{0,0,0}{71,237}\\
Unseen test & 67 & \textcolor[rgb]{0,0,0}{7,355} & \textcolor[rgb]{0,0,0}{63,882} & \textcolor[rgb]{0,0,0}{71,237}\\
\hline
\end{tabular}
\end{table}

\subsection{Noisy LA Dataset}
The bona fide and spoofed utterances of the LA dataset of ASVspoof 2019 are under clean conditions. However, the audio segments of our SceneFake dataset are under noisy conditions. So we also design a noisy LA dataset to evaluate whether the manipulated utterances can be detected reliably with existing audio fake detection models under the noisy conditions.

The noisy LA dataset is generated based upon the LA dataset of ASVspoof 2019. The bona fide and spoofed utterances from the LA dataset are randomly mixed with scenes from the acoustic scene dataset in DCASE 2022 challenge as shown in Table~\ref{tab:scene}.
The utterances of training, development and seen test set in the noisy LA dataset are generated based upon that of training, development and test set from the LA dataset, respectively. The utterances in these three sets are generated by using six scenes: Airport, Bus, Park, Public, Shopping, Station. The voices of unseen test set are simulated with four scenes: Metro, Pedestrian, Street, Tram.
The acoustic scenes are randomly sampled to mix with the bona fide and spoofed utterances at 6 different SNRs \blackf{each:} -5dB, 0dB, 5dB, 10dB, 15dB and 20dB.

\blackf{We make the total number of utterances of noisy LA dataset be equal to that of the LA dataset of ASVspoof 2019} so that the performance can be compared fairly.
Table~\ref{tab:la-noisy} illustrates the statistic distribution of the noisy LA dataset, \blackf{where \#Speakers, \#Genuine, \#Spoofed, and \#Total denote the number of speakers, genuine utterances, spoofed utterances, and all utterances in noisy LA dataset of ASVspoof 2019.}

\subsection{Experimental Settings}

Motivated by the baselines of \blackf{the} ASVspoof 2021~\cite{2021ASVspoof} and ADD 2022~\cite{Yi2022ADD} challenges, we employ Gaussian mixture model (GMM), light convolutional neural network (LCNN) and RawNet2 to train fake audio detection models as our baselines.
We use the officially released source code~\footnote{http://github.com/asvspoof-challenge/2021} with minor modification to build GMM, LCNN and RawNet2 models. Linear frequency cepstral coefficients (LFCCs) are used as the input features of GMM and LCNN based models. The input features of RawNet2 models are raw audio waveforms. The classifiers of the models are standard 2-class discriminators. The output labels are ``\textit{fake}'' and ``\textit{real}''. The source codes of three baseline models are \blackf{publicly} available~\footnote{https://github.com/ADDchallenge/SceneFake}.

GMM: features are extracted using a 30-ms sliding window with a 15-ms shift, a 1024-point Fourier transform and
70 filters. \blackf{The} LFCC features consist of 19 static cepstral coefficients appended by energy (C0 or $0^{th}$ cepstral coefficients) delta and delta-delta coefficients.

LCNN: LFCC features are extracted with a 20-ms sliding window with a 10-ms shift, a 1024-point Fourier transform and 70 filters. The features are comprised of 19 static cepstral coefficients plus energy, delta and delta-delta coefficients. The features are extracted with a 20-ms sliding window with a 10-ms shift. The architecture of the LCNN based models~\cite{Wu2020LCNN} \blackf{incorporates} average pooling and LSTM layers~\cite{2021ASVspoof}.

RawNet2: \blackf{The RawNet2~\cite{Jung2020Rawnet2} is a fully end-to-end model, which operates directly upon raw audio waveforms upon raw audio waveforms as input features}. It \blackf{consists of} one fixed bank of sinc filters and six residual blocks. A gated recurrent units and fully connected layers prior followed by the output layer which has two labels: real or fake.

We only utilize the respective training data to train the models and use the respective development data to optimize the model. The development sets are employed to choose better models and hyper parameters. The training stops if there \blackf{is} only a little improvement between two adjacent epochs. We also do not utilize any kind of data augmentation technology.
The fake audio detection models are evaluated in terms of EER and \blackf{t-DCF} on test sets.

\begin{table*}
\caption{The results of \blackf{the fake utterances using different speech enhancement models} in terms of PESQ on our SceneFake dataset. ``Avg.'' denotes the average PESQ of the fake utterances using all speech enhancement models on the \blackf{corresponding} sets. \blackf{``BeforeSE'' denotes the results of the original utterances of fakes ones, which are not enhanced by speech enhancement models. The BeforeSE utterances are mixed with acoustic scenes.} ``Total'' \blackf{is referred to as} the performance of the models at all \blackf{SNR$_{fake}$}.}
\label{tab:pesq-se}
\centering
\begin{tabular}{|c||c|ccccccc|}
\hline
\multirow{2}{*}{\makecell[c]{Set}} &
\multirow{2}{*}{Models} &
\multicolumn{7}{c|}{PESQ}\\
\cline{3-9}
 &  & -5dB &  0dB & 5dB  & 10dB & 15dB  &  20dB & Total\\
\hline
 \multirow{4}{*}{\makecell[c]{Training}}
 & SSub & 1.23  & 1.88  & 2.08  & 2.35  & 3.15  & 2.86  & 2.26 \\
 & MMSE & 1.28  & 1.89  & 2.10  & 2.36  & 3.18  & 2.88  & 2.28\\
 & Wiener& 1.37  & 1.98  & 2.26  & 2.44  & 3.28  & 3.06  & 2.41 \\
 & FullSubNet& 1.73  & 2.27  & 2.68  & 2.84  & 3.25  & 3.38  & 2.75 \\
 &Avg.& 1.40  & 2.01  & 2.28  & 2.50  & 3.22  & 3.05  & 2.43 \\
 \cline{2-9}
  & \blackf{BeforeSE} &  \blackf{1.07}  & \blackf{1.43}   & \blackf{1.69}   & \blackf{1.95}   & \blackf{2.21}   & \blackf{2.13}   & \blackf{1.92}  \\
\hline
\multirow{4}{*}{\makecell[c]{Development}}
 & SSub & 1.26  & 1.88  & 2.11  & 2.38  & 3.16  & 2.88  & 2.27 \\
 & MMSE & 1.27  & 1.88  & 2.15  & 2.44  & 3.18  & 2.91  & 2.32 \\
 & Wiener& 1.32  & 1.95  & 2.26  & 2.45  & 3.25  & 3.05  & 2.45 \\
 &FullSubNet& 1.75  & 2.28  & 2.72  & 2.86  & 3.27  & 3.38  & 2.77 \\
 &Avg.& 1.40  & 2.00  & 2.31  & 2.53  & 3.22  & 3.06  & 2.45 \\
 \cline{2-9}
  & \blackf{BeforeSE} &  \blackf{1.07}  & \blackf{1.46}   & \blackf{1.72}   &  \blackf{1.92}  &  \blackf{2.15}  & \blackf{2.08}   & \blackf{1.95}  \\
\hline
 \multirow{4}{*}{\makecell[c]{Seen test}}
 & SSub & 1.26  & 1.87  & 2.05  & 2.37  & 3.17  & 2.81  & 2.24 \\
 & MMSE & 1.30  & 1.88  & 2.09  & 2.37  & 3.20  & 2.84  & 2.28\\
 & Wiener& 1.38  & 1.96  & 2.27  & 2.49  & 3.26  & 3.06  & 2.46  \\
 &FullSubNet& 1.78  & 2.27  & 2.67  & 2.85  & 3.28  & 3.39  & 2.77 \\
 &Avg.& 1.43  & 2.00  & 2.27  & 2.52  & 3.23  & 3.03  & 2.44 \\
 \cline{2-9}
 & \blackf{BeforeSE} & \blackf{1.05}   & \blackf{1.45}  & \blackf{1.68}   & \blackf{1.97}   &  \blackf{2.18}  &  \blackf{2.10}  & \blackf{1.91}  \\
\hline
\multirow{3}{*}{\makecell[c]{Unseen test}}
 & WaveU-Net & 1.56  & 2.05  & 2.39  & 2.65  & 3.31  & 3.17  & 2.44\\
 & GCRN & 1.63  & 2.09  & 2.47  & 2.61  & 3.19  & 3.26  & 2.51 \\
 &Avg.& 1.60  & 2.07  & 2.43  & 2.63  & 3.25  & 3.22  & 2.48 \\
 \cline{2-9}
 & \blackf{BeforeSE} & \blackf{1.12}   &  \blackf{1.48}  & \blackf{1.70}   & \blackf{1.95}   &  \blackf{2.17}  & \blackf{2.10}   & \blackf{1.94}  \\
\hline
\end{tabular}
\end{table*}

\begin{table*}
\caption{The results of \blackf{the fake utterances using different speech enhancement models} in terms of STOI on our SceneFake dataset. ``Avg.'' denotes the average STOI of the fake utterances using all speech enhancement models on the \blackf{corresponding} sets. \blackf{``BeforeSE'' denotes the results of the original utterances of fakes ones, which are not enhanced by speech enhancement models. The BeforeSE utterances are mixed with acoustic scenes.} ``Total'' \blackf{ is referred to as} the performance of the models at all \blackf{SNR$_{fake}$}.}
\label{tab:stoi-se}
\centering
\begin{tabular}{|c||c|ccccccc|}
\hline
\multirow{2}{*}{\makecell[c]{Set}} &
\multirow{2}{*}{Models} &
\multicolumn{7}{c|}{STOI}\\
\cline{3-9}
 &  & -5dB &  0dB & 5dB  & 10dB & 15dB  &  20dB & Total\\
\hline
 \multirow{4}{*}{\makecell[c]{Training}}
 & SSub & 0.58 & 0.62 & 0.64 & 0.83 & 0.86 & 0.87 & 0.74\\
 & MMSE &  0.60 & 0.65 & 0.65 & 0.82 & 0.84 & 0.89 & 0.75\\
 & Wiener&  0.62 & 0.66 & 0.67 & 0.82 & 0.86 & 0.89 & 0.75 \\
 & FullSubNet& 0.68 & 0.72 & 0.78 & 0.91 & 0.94 & 0.95 & 0.84\\
 &Avg.&  0.62 & 0.66 & 0.69 & 0.85 & 0.88 & 0.90 & 0.77 \\
 \cline{2-9}
  & \blackf{BeforeSE} & \blackf{0.41}   & \blackf{0.47}   & \blackf{0.49}   &  \blackf{0.58}  & \blackf{0.59}   & \blackf{0.62}   & \blackf{0.52}  \\
\hline
\multirow{4}{*}{\makecell[c]{Development}}
& SSub &  0.60 & 0.62 & 0.65 & 0.83 & 0.86 & 0.87 & 0.74 \\
 & MMSE & 0.61 & 0.64 & 0.64 & 0.83 & 0.85 & 0.88 & 0.74\\
 & Wiener&  0.61 & 0.65 & 0.65 & 0.82 & 0.83 & 0.87 & 0.75 \\
 &FullSubNet& 0.69 & 0.72 & 0.80 & 0.91 & 0.95 & 0.95 & 0.85  \\
 &Avg. & 0.63 & 0.66 & 0.69 & 0.85 & 0.87 & 0.89 & 0.77 \\
 \cline{2-9}
 & \blackf{BeforeSE} & \blackf{0.43}   & \blackf{0.43}   & \blackf{0.46}   & \blackf{0.55}   & \blackf{0.57}   &\blackf{0.61}    &\blackf{0.51}   \\
\hline
 \multirow{4}{*}{\makecell[c]{Seen test}}
 & SSub & 0.61 & 0.61 & 0.65 & 0.81 & 0.84 & 0.88 & 0.75\\
 & MMSE &  0.62 & 0.65 & 0.65 & 0.83 & 0.85 & 0.89 & 0.76\\
 & Wiener&  0.62 & 0.66 & 0.67 & 0.82 & 0.86 & 0.88 & 0.76 \\
 &FullSubNet&  0.68 & 0.70 & 0.79 & 0.92 & 0.94 & 0.96 & 0.84 \\
 &Avg.& 0.63 & 0.66 & 0.69 & 0.85 & 0.87 & 0.90 & 0.78 \\
 \cline{2-9}
 & \blackf{BeforeSE} & \blackf{0.41}   & \blackf{0.46}   & \blackf{0.49}   & \blackf{0.57}   &  \blackf{0.59}  & \blackf{0.63}   & \blackf{0.53}  \\
\hline
\multirow{2}{*}{\makecell[c]{Unseen test}}
& WaveU-Net & 0.65 & 0.68 & 0.72 & 0.83 & 0.89 & 0.91 & 0.77 \\
 & GCRN & 0.66 & 0.68 & 0.74 & 0.86 & 0.89 & 0.92 & 0.79 \\
 &Avg.& 0.66 & 0.68 & 0.73 & 0.85 & 0.89 & 0.92 & 0.78 \\
 \cline{2-9}
 & \blackf{BeforeSE} &  \blackf{0.40}  & \blackf{0.48}   & \blackf{0.51}   & \blackf{0.56}   & \blackf{0.58}   & \blackf{0.61}   & \blackf{0.53}  \\
\hline
\end{tabular}
\end{table*}

\subsection{Performance of Speech Enhancement models}
The scene manipulated audio is tampered \blackf{with} speech enhancement technologies in this work. We employ open source speech enhancement models to estimate an enhanced utterance by removing the original scene of a real utterance. A fake utterance is generated by mixing the enhanced utterance with another scene. We evaluate the performance of various speech enhancement models in this section.

The traditional speech enhancement models used in this work are SSub, MMSE and Wiener, source codes of which are public available~\footnote{https://github.com/fchest/traditional-speech-enhancement}.
The neural network based models are FullSubNet~\footnote{https://github.com/haoxiangsnr/fullsubnet}, WaveU-Net~\footnote{https://github.com/haoxiangsnr/Wave-WaveU-Net-for-Speech-Enhancement} and GCRN~\footnote{https://github.com/JupiterEthan/GCRN-complex}, which are trained \blackf{using clean and noisy WSJ0-SI84 datasets~\cite{Two2021Li}.}
The results of all speech enhancement models on our SceneFake dataset are \blackf{reported in terms of perceptual evaluation of speech quality (PESQ)~\cite{Rix2002Pesq}} and short-time objective intelligibility (STOI)~\cite{Taal2010ASO}. PESQ is able to predict subjective quality of the enhanced speech in a very wide range of conditions. STOI is an objective intelligibility measure. The results of the models at different \blackf{SNR$_{fake}$} on the SceneFake dataset are showed in \blackf{Tables~\ref{tab:pesq-se} and ~\ref{tab:stoi-se}}. ``Avg.'' denotes the average PESQ or STOI of different models on the \blackf{corresponding} sets and ``Total'' \blackf{refers to} the performance of the models at all \blackf{SNR$_{fake}$}. \blackf{``BeforeSE'' denotes the results on unenhanced fake utterances, i.e., those whose scene has not been removed through speech enhancement, being mixed with new acoustic scenes.}

For the training, development and seen test sets, the results show that the SSub model obtains the worst performance and the FullSubNet achieves the best results. The MMSE model outperforms the SSub model but underperforms the Wiener model. The results also demonstrate that all the models obtain the worst PESQ and STOI at -5dB. The SSub, MMSE and Wiener models achieve the best PESQ at 15dB but obtain the best STOI at 20dB. The FullSubNet model obtains the best PESQ and STOI both at 20dB.

\begin{table*}[!t]
\caption{EERs (\%) and t-DCF of the models trained with the training data of LA dataset in ASVspoof 2019 (under clean conditions).}
\centering
\label{tab:la-train}
\setlength{\tabcolsep}{0.8mm}{
\begin{tabular}{|c||cccccccccc|}
\hline
\multirow{3}{*}{Models} &
\multicolumn{10}{c|}{Test set} \\
\cline{2-11}
& \multicolumn{2}{c}{LA} & \multicolumn{2}{c}{Noisy LA seen}  & \multicolumn{2}{c}{Noisy LA unseen} & \multicolumn{2}{c}{Our seen} & \multicolumn{2}{c|}{Our unseen} \\
\cline{2-11}
& EER & \blackf{t-DCF} & EER &\blackf{t-DCF}& EER & \blackf{t-DCF}& EER & \blackf{t-DCF}& EER & \blackf{t-DCF}  \\
\hline
GMM & 8.72 & \blackf{0.214} & 15.82 &\blackf{0.621} & 13.69 &\blackf{0.438} & 65.74&\blackf{1.000} & 51.88&\blackf{1.000}  \\
LCNN & 4.75 &\blackf{0.108}& 11.24 &\blackf{0.405} & 9.82 &\blackf{0.402} & 54.51 &\blackf{1.000}& 42.30 &\blackf{0.984} \\
Rawnet2 & 5.46 &\blackf{0.117}& 13.75 &\blackf{0.418} & 10.44 &\blackf{0.397} & 53.82& \blackf{0.993}  & 44.16 &\blackf{0.988} \\
Avg. & 6.31&\blackf{0.173} & 13.60 & \blackf{0.445} & 11.32 & \blackf{0.416} & 58.02 &\blackf{1.000} & 46.11 &\blackf{0.997}\\
\hline
\end{tabular}}
\end{table*}

\begin{table*}[!t]
\caption{EERs (\%) and t-DCF of the models trained with the noisy training data of noisy LA dataset simulated upon the LA dataset of ASVspoof 2019.}
\centering
\label{tab:noisyla-train}
\begin{tabular}{|c||cccccccc|}
\hline
\multirow{3}{*}{Models} &
\multicolumn{8}{c|}{Test set} \\
\cline{2-9}
& \multicolumn{2}{c}{Noisy LA seen} & \multicolumn{2}{c}{Noisy LA unseen} & \multicolumn{2}{c}{Our seen} & \multicolumn{2}{c|}{Our unseen} \\
\cline{2-9}
& EER & \blackf{t-DCF}& EER & \blackf{t-DCF}& EER & \blackf{t-DCF}& EER & \blackf{t-DCF}  \\
\hline
GMM &  7.93 &\blackf{0.340} &  7.49 &\blackf{0.337} & 44.94 &\blackf{0.999} &  36.71 &\blackf{0.914} \\
LCNN &  2.86 &\blackf{0.247}  &  2.36 &\blackf{0.251}  & 35.85&\blackf{0.835} &  26.94&\blackf{0.738} \\
Rawnet2 &  5.22 &\blackf{0.284} & 4.78 &\blackf{0.276} & 37.86 &\blackf{0.870} &  28.12&\blackf{0.766} \\
Avg. & 5.34 &\blackf{0.292} & 4.88 &\blackf{0.282} & 39.55&\blackf{0.874} & 30.59&\blackf{0.821} \\
\hline
\end{tabular}
\end{table*}

For the unseen test \blackf{set}, the results demonstrate that the WaveU-Net model underperforms the GCRN model. The results also show that the WaveU-Net and GCRN model both obtain the lowest PESQ and STOI at -5dB \blackf{and} the highest PESQ and STOI at 20dB.

\subsection{Performance of Baseline Models of ASVspoof 2019}
We conduct several groups of experiments to evaluate the performance of baseline models of \blackf{the} LA task in ASVspoof 2019.

In the first group of experiments, we want to know whether the manipulated utterances can be detected effectively by the baseline models of ASVspoof 2019. We report the results of the baseline models trained using the training set of LA dataset in ASVspoof 2019. The results are reported on seen and unseen test sets of our SceneFake dataset. We also compare the results to that on test set of LA dataset and seen and unseen test sets of noisy LA dataset. The results are listed in Table~\ref{tab:la-train}.
The results show that all the detection models obtain the best performance on the test set of LA dataset \blackf{and achieve} worse performance on test sets of noisy LA dataset. Futhermore, their performance on \blackf{two test sets} of SceneFake dataset \blackf{degrades} significantly. The average \blackf{EER and t-DCF of the models on the two test sets} of our SceneFake dataset are absolutely higher than that on the test sets of LA dataset by up to \blackf{51.71\% and 0.872.} The results on the unseen test set of noisy LA are better than that on the seen test set. This is because that there are only four kinds of scenes contained in the unseen test set but six kinds of scenes involved in the seen test set. The results on the unseen test sets of \textcolor[rgb]{0,0,0}{the} SceneFake dataset are better than that on the seen test set.

\begin{table*}[!t]
	\caption{EERs (\%) and t-DCF of the models trained with the training data of our SceneFake dataset. ``Avg.'' denotes the average EER \blackf{or t-DCF} of different baseline models on the \blackf{corresponding} test sets.}
	\centering
	\label{tab:SceneFake-train}
	\begin{tabular}{|c||c|cccc|}
		\hline
		\multirow{2}{*}{Training set} & \multirow{2}{*}{Models} & \multicolumn{2}{c}{Our seen test} & \multicolumn{2}{c|}{Our unseen test} \\
		\cline{3-6}
		& & EER & \blackf{t-DCF}& EER & \blackf{t-DCF} \\
		\cline{2-6}
		\multirow{4}{*}{Our SceneFake} & GMM &  4.59 & \blackf{0.282} &  23.21 & \blackf{0.695} \\
		& LCNN & 0.78 & \blackf{0.148} &  16.72 & \blackf{0.618}\\
		& Rawnet2 & 0.85 & \blackf{0.184} &  15.31 & \blackf{0.625}\\
		& Avg. & 2.07 & \blackf{0.188} & 18.41 & \blackf{0.633}\\
		\hline
	\end{tabular}
\end{table*}

\begin{table*}
\caption{EERs (\%) of the models on our test set which \blackf{has been manipulated} with different speech enhancement models.}
\label{tab:SceneFake-train-se}
\centering
\begin{tabular}{|c||cccc|cc|}
\hline
\multirow{2}{*}{Models} &
\multicolumn{4}{c|}{Our seen test}&
\multicolumn{2}{c|}{Our unseen test}\\
\cline{2-7}
 & SSub & MMSE & Wiener&  FullSubNet &  WaveU-Net  & GCRN \\
\hline
GMM & 5.85 & 5.37 & 4.17& 3.93  & 25.26 & 23.84\\
LCNN  & 1.49 & 0.52 & 0.11& 0.07 & 17.61 & 16.23 \\
Rawnet2 & 1.22 & 0.63 & 0.16 & 0.13  & 15.51 & 16.84 \\
Avg.  & 2.85  & 2.17  & 1.48 & 1.38  & 19.46  & 18.97  \\
\hline
\end{tabular}
\end{table*}

The utterances of the LA dataset of ASVspoof 2019 are under clean conditions. However, the audio segments of our SceneFake dataset are under noisy conditions. In order to compare fairly, we assess the performance of the detection models under the noisy conditions in the second group of experiments. The detection models \blackf{are} trained with the training set of noisy LA dataset \blackf{and} evaluated on \blackf{the} seen and unseen test sets of our SceneFake dataset. \blackf{In addition}, the results are compared to that on seen and unseen test sets of noisy LA dataset. The results \blackf{reported in Table~\ref{tab:noisyla-train}, demonstrate} that all the detection models obtain better performance on the test sets of noisy LA dataset. The average \blackf{EER and t-DCF} of the models on \blackf{two test sets} of our SceneFake dataset are absolutely higher than that on the test sets of noisy LA dataset by up to \blackf{34.21\% and 0.582, respectively}.

\begin{table*}
\caption{\blackf{t-DCF of the models on our test set which has been manipulated with different speech enhancement models.}}
\label{tab:SceneFake-train-se-tdcf}
\centering
\begin{tabular}{|c||cccc|cc|}
\hline
\multirow{2}{*}{\blackf{Models}} &
\multicolumn{4}{c|}{\blackf{Our seen test}}&
\multicolumn{2}{c|}{\blackf{Our unseen test}}\\
\cline{2-7}
 & \blackf{SSub} & \blackf{MMSE} & \blackf{Wiener}&  \blackf{FullSubNet} &  \blackf{WaveU-Net}  & \blackf{GCRN} \\
\hline
\blackf{GMM} & \blackf{0.306} & \blackf{0.294} & \blackf{0.285}& \blackf{0.297}& \blackf{0.682}& \blackf{0.694} \\
\blackf{LCNN}  & \blackf{0.164} & \blackf{0.137} & \blackf{0.088}& \blackf{0.092}& \blackf{0.596}& \blackf{0.613}\\
\blackf{Rawnet2} & \blackf{0.165} & \blackf{0.142} & \blackf{0.095}& \blackf{0.092}& \blackf{0.603}& \blackf{0.619} \\
\blackf{Avg.} & \blackf{0.196} & \blackf{0.188} & \blackf{0.175}& \blackf{0.172}& \blackf{0.628}& \blackf{0.636}  \\
\hline
\end{tabular}
\end{table*}

Therefore, although the baseline model of the ASVspoof 2019 obtain \blackf{the lowest EER and t-DCF}, it is difficult for them to correctly detect the scene fake utterances.
Moreover, it is still hard for the baseline models to distinguish the real from manipulated audios, even when the models \blackf{are} trained using the data under similar noisy environments.

\subsection{Baselines on Our SceneFake dataset}

We conduct experiments to evaluate the performance of baseline models on our SceneFake dataset in this section.
The three baseline models are trained with the training set of our SceneFake dataset. ``Avg.'' denotes the average EER \blackf{or t-DCF} of different baseline models on the \blackf{corresponding} test sets.

\blackf{The EER and t-DCF} of our baselines are reported on seen and unseen test sets of our SceneFake dataset, which are listed in Table~\ref{tab:SceneFake-train}.
\blackf{The results show that the models perform better on the seen test set than on the unseen test set.} This is because the acoustic distribution of the seen test set is similar to that of the training set. But there exists \blackf{a} mismatch between the training set and the unseen test \blackf{set}.

\begin{table*}
\caption{EERs (\%) of the models on our test sets of the SceneFake dataset at different SNRs. ``Avg.'' denotes the average EER of different baseline models on the corresponding test sets. ``Total'' \blackf{ is referred to as} the performance of the baseline models at all SNRs.}
\label{tab:SceneFake-train-snr}
\centering
\begin{tabular}{|c||ccccccc|}
\hline
\multirow{1}{*}{\makecell[c]{Test sets}} &
\multirow{1}{*}{Models} &
 -5dB &  0dB & 5dB  & 10dB & 15dB  &  20dB  \\
\hline
 \multirow{3}{*}{Our seen test} & GMM & 4.12 & 4.66 & 5.78 & 4.92 & 5.32 & 5.98  \\
 & LCNN & 0.00 & 0.05 & 0.42 & 0.05 & 0.41 & 0.79  \\
 & Rawnet2 & 0.23 & 0.17 & 0.72 & 0.14 & 0.77 & 0.93  \\
  & Avg. & 1.45 & 1.63 & 2.31 & 1.70 & 2.17 & 2.57  \\
\cline{1-8}
 \multirow{3}{*}{Our unseen test} & GMM & 31.38 & 3.91 & 4.58 & 4.79 & 4.77 & 21.12  \\
  & LCNN & 26.55 & 0.17 & 0.34 & 2.62 & 1.58 & 14.32 \\
  & Rawnet2 & 23.15 & 0.35 & 0.55 & 2.81 & 1.41 & 15.88 \\
 & Avg. & 27.03  & 1.48  & 1.82  & 3.41  & 2.59  & 17.11 \\
\hline
\end{tabular}
\end{table*}

\blackf{The EER and t-DCF} of our baselines using different speech enhancement models are reported on \blackf{the seen and unseen test sets} as shown in Tables~\ref{tab:SceneFake-train-se} and~\ref{tab:SceneFake-train-se-tdcf}.
On the test set, the results show that the better performance of the speech enhancement model the lower average EER of the detection model we obtain. The detection models with ``\textit{FullSubNet}'' speech enhancement model obtain the best performance. But the detection models with ``\textit{SSub}'' speech enhancement model achieve the worst results.
On the unseen test set, the average EER of the detection models with ``\textit{WaveU-Net}'' speech enhancement model \blackf{is higher than} that with ``\textit{GCRN}'' speech enhancement model\blackf{, while the average t-DCF of the detection models with ``\textit{WaveU-Net}'' speech enhancement model is lower than that with ``\textit{GCRN}'' speech enhancement model.}

\begin{table*}
	\caption{\blackf{t-DCF of the models on our test sets of the SceneFake dataset at different SNRs. ``Avg.'' denotes the average t-DCF of different baseline models on the corresponding test sets. ``Total'' \blackf{ is referred to as} the performance of the baseline models at all SNRs.}}
	\label{tab:SceneFake-train-snr-tdcf}
	\centering
	\begin{tabular}{|c||ccccccc|}
		\hline
		\multirow{1}{*}{\makecell[c]{\blackf{Test sets}}} &
		\multirow{1}{*}{\blackf{Models}} &
		\blackf{-5dB} &  \blackf{0dB} & \blackf{5dB}  & \blackf{10dB} & \blackf{15dB}  &  \blackf{20dB}  \\
		\hline
		\multirow{4}{*}{\blackf{Our seen test}} & \blackf{GMM} & \blackf{0.291} & \blackf{0.287}& \blackf{0.301}& \blackf{0.299}& \blackf{0.295}& \blackf{0.302} \\
		& \blackf{LCNN} & \blackf{0.000} & \blackf{0.038}& \blackf{0.132}& \blackf{0.049}& \blackf{0.129}& \blackf{0.145}  \\
		& \blackf{Rawnet2} & \blackf{0.108} & \blackf{0.096}& \blackf{0.153}& \blackf{0.090}& \blackf{0.149}& \blackf{0.161}  \\
		& \blackf{Avg.} & \blackf{0.153} & \blackf{0.159}& \blackf{0.192}& \blackf{0.165}& \blackf{0.184}& \blackf{0.193}  \\
		\cline{1-8}
		\multirow{4}{*}{\blackf{Our unseen test}} & \blackf{GMM} & \blackf{0.826} & \blackf{0.259}& \blackf{0.266}& \blackf{0.266}& \blackf{0.270}& \blackf{0.726} \\
		& \blackf{LCNN} & \blackf{0.762} & \blackf{0.098}& \blackf{0.125}& \blackf{0.198}& \blackf{0.158}& \blackf{0.529}  \\
		& \blackf{Rawnet2} & \blackf{0.758} & \blackf{0.120}& \blackf{0.134}& \blackf{0.206}& \blackf{0.153}& \blackf{0.547}  \\
		& \blackf{Avg.} & \blackf{0.771} & \blackf{0.156}& \blackf{0.174}& \blackf{0.242}& \blackf{0.218}& \blackf{0.616}  \\
		\hline
	\end{tabular}
\end{table*}

\blackf{The EER and t-DCF} of baseline models on \blackf{the seen and \blackf{unseen test set}} at different SNRs are listed in Tables~\ref{tab:SceneFake-train-snr} and ~\ref{tab:SceneFake-train-snr-tdcf}.
The results on the seen test set show that the baseline models achieve the best performance at -5dB but obtain the worst result at 20dB.
However, the results on \blackf{the unseen test set} show that the baseline models achieve the worst performance at -5dB but obtain the best results at 0dB. The results on the \blackf{unseen test set} also show that the models achieve \blackf{noticeably} higher \blackf{EER and t-DCF} at -5dB and 20dB than that at other SNRs.
\blackf{A} possible reason may be that the baseline models try to distinguish the real utterance from the fake one mainly using acoustic scene information and distorted manipulation traces.

\begin{figure*}[tb]
\hfill
\begin{minipage}[b]{1.0\linewidth}
  \centering
  \centerline{\includegraphics[width=12.5cm,height=7.1cm]{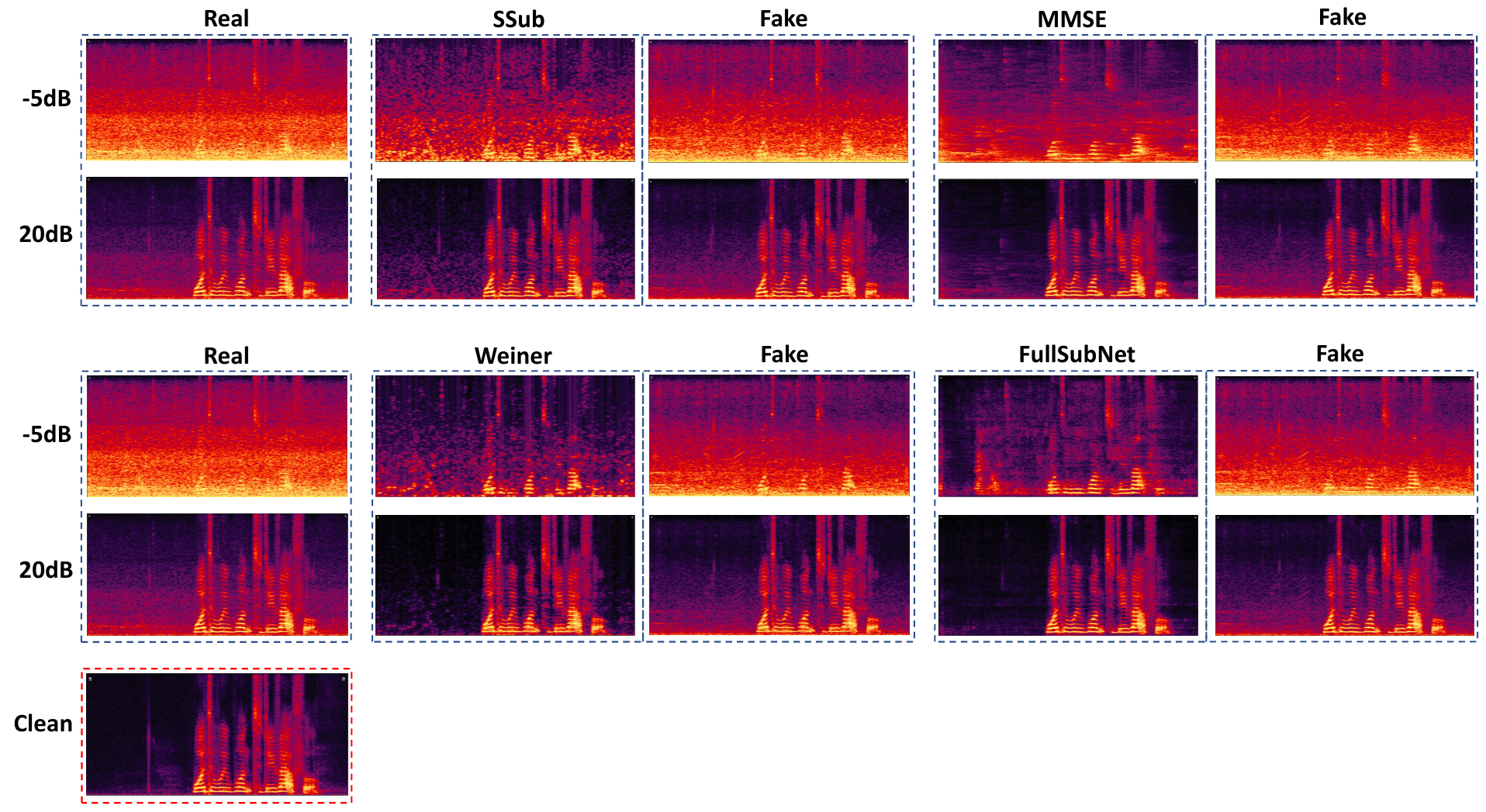}}
\end{minipage}
\caption{Spectrogram examples of utterances at -5dB and 20dB in the seen test set. ``Clean'' denotes the clean utterance. ``Real'' denotes the real utterances simulated by adding acoustic scene ``\textit{Airport}'' to the clean utterance. ``SSub'', ``MMSE'', ``Weiner'', ``FullSubNet'' denote the respective enhanced utterance with the respective speech enhancement model. ``Fake'' denotes the respective fake utterance generated by adding scenes ``\textit{Public square}'' to the respective enhanced speech.}
\label{fig:analysis-seen}
\end{figure*}

\begin{figure*}[tb]
\hfill
\begin{minipage}[b]{1.0\linewidth}
  \centering
  \centerline{\includegraphics[width=12.5cm,height=2.9cm]{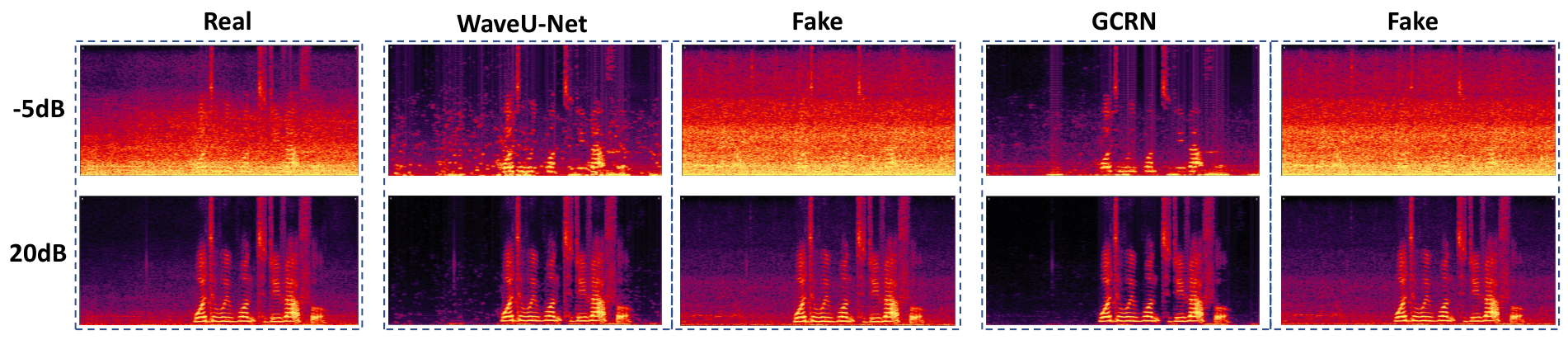}}
\end{minipage}
\caption{Spectrogram examples of utterances at -5dB and 20dB in the unseen test set. ``Real'' denotes the real utterances simulated by adding acoustic scene ``\textit{Metro}'' to the clean utterance that is the same one in Figure \ref{fig:analysis-seen}.. ``WaveU-Net'' and  ``GCRN'' denote the respective enhanced utterance with the respective speech enhancement model. ``Fake'' denotes the respective fake utterance generated by adding scenes ``\textit{Street}'' to the respective enhanced speech. }
\label{fig:analysis-unseen}
\end{figure*}

Acoustic scenes of real and fake \blackf{utterances} are the dominated signals at -5dB as shown in \textcolor[rgb]{0,0,0}{Figures~\ref{fig:analysis-seen} and \ref{fig:analysis-unseen}}. In addition, manipulation traces of fake audios are the strongest at -5dB, which are brought by speech enhancement models. Signals of real and fake audios are both significantly distorted by adding scene noises at -5dB.
\textcolor[rgb]{0,0,0}{However, the actual speech content becomes more dominant as the SNR increases, and the manipulation traces from speech enhancement models are the weakest at 20 dB, as shown in Figures~\ref{fig:analysis-seen} and \ref{fig:analysis-unseen}.} Also, \blackf{the addition of scene noises does not significantly distort speech signals in real and fake audios.} Therefore, the acoustic scenes are strong and the distorted manipulation traces are obvious at -5dB. But the acoustic scenes are weak and the distorted manipulation traces are slight at 20dB.

When the data distribution of the seen test set matches the training set, it is very easy to distinguish the real one from the fake by using known strong acoustic scenes and known obvious distorted manipulation traces at -5dB. The baseline model achieves the best performance at -5dB on the seen test set. But it is still hard to distinguish the real one from the fake by using known slight acoustic scenes and known weak distorted manipulation traces at 20dB even when the data distribution of the seen test set matches the training set. The baseline model achieves the worst result at 20dB on the seen test set.

However, \blackf{there exists a significant difference between the data distribution of the unseen test set and that of the training set,} which is brought by unseen acoustic scenes and unseen speech enhancement models. Thus it is difficult to distinguish the real one from the fake by using unknown strong acoustic scenes and unknown obvious distorted manipulation traces at -5dB. \blackf{In addition}, it is hard to distinguish the real one from the fake by using unknown slight acoustic scenes and unknown weak distorted manipulation traces at 20dB.

Although the models \blackf{perform well} on the seen test set, their performance \blackf{is} very poor on the unseen test set. Therefore, it is still challenging to effectively detect the scene manipulated fake utterances.

\section{Discussions}
\label{sec:dis}

Some findings and observations of this work are summarized in this section. Furthermore, we discuss some limitations of our work.
Both the findings and the limitations are suggested to be researched further.

\subsection{Findings}
The above benchmark results show some interesting observations and findings.

\begin{itemize}
\item{Manipulated utterances can not be detected reliably with existing \blackf{LA} baseline models of ASVSpoof 2019.} Although the existing fake audio detection systems obtain low EER \blackf{and t-DCF} on the ASVspoof dataset, \blackf{they face challenges in detecting} detecting scene manipulated audios correctly. Moreover, it is still hard for the existing the LA baseline models of ASVSpoof 2019 to distinguish the real from manipulated audios, even when the models trained using the data under similar noisy environments.

\item{Scene manipulation audio detection task is still challenging.} Although the models trained using the training set of \textcolor[rgb]{0,0,0}{the} SceneFake achieve a good performance on the seen test set, \blackf{they still perform poorly on the unseen test set.} Therefore, it is challenging to detect reliably unknown scene fake fake utterances.

\item{The detection models achieve the best performance on seen test but obtain the worst results on \blackf{unseen test set} at -5dB.} When the data distribution of the seen test set matches the training set, it is very easy to discriminate the real from the fake by using known strong acoustic scenes and known obvious distorted manipulation traces at -5dB. However, it is hard to distinguish the real one from the fake at -5dB when \blackf{a} mismatch exists between the unseen test set and the training set.

\item{The detection models obtain poor performance both on seen and unseen test sets at 20dB.} It is difficult to tell the difference between \blackf{real and the fake utterances} using unknown slight acoustic scenes or unknown weak distorted manipulation traces at 20dB.
\blackf{In addition}, it is still hard to discriminate the real one from the fake at 20dB even when the data distribution of the seen test set matches the training set.

\end{itemize}

\subsection{\textcolor[rgb]{0,0,0}{Limitations}}
Although we have designed an initial dataset and conducted benchmarking experiments for audio scene manipulation detection, there are still some limitations which \blackf{should be addressed} in future work.
\begin{itemize}
\item{\textcolor[rgb]{0,0,0}{Collecting utterances under realistic conditions:}} The utterances of the current SceneFake dataset are simulated data generated by mixing clean utterances with different scenes. Such emulations do not quite match with the real utterances recorded in real conditions. For instance, the linguistic content of the audio may not match up with the scene. The mismatch between the two will not happen in real recordings. Moreover, the real conditions of the utterances may be even worse and vary \blackf{more greatly} than the simulated conditions. In order to \blackf{assess} manipulation detection methods in practical applications, the utterances with a variety of scenes \blackf{need} to be collected through realistic environment conditions, such as social media platforms.

\item{\textcolor[rgb]{0,0,0}{More diverse manipulation types:}} Our dataset here involves ten kinds of acoustic scenes and six sorts of speech enhancement methods \blackf{, while} the acoustic scenes in the recordings are more diverse and complex in real-life scenarios. \blackf{In addition}, more kinds of speech enhancement technologies are utilized to manipulated the original audio. It is crucial to \blackf{take into consideration more approaches to generate scene fake audio, so that the proposed dataset may be more appropriate for real scenarios.}

\item{\textcolor[rgb]{0,0,0}{Considering cross language scenarios:}} The current work only designs an \textcolor[rgb]{0,0,0}{English-language manipulated} audio dataset and conduct benchmark experiments on \blackf{the said} dataset. It may make the detection methods language dependent. But it is important to evaluate the performance of fake detection models in the cross language scenario and for code-switching between different languages. In order to make fake detection systems more \blackf{robust} for other languages, it is necessary to design and develop multi-lingual datasets and \blackf{language independent forgery countermeasures}.

\item{\textcolor[rgb]{0,0,0}{Robustness and generalization of detection methods:}} The work here aims to provide benchmark results on the SceneFake dataset for future research. The GMM, LCNN and RawNet2 models employed in this work are the baseline models in ASVspoof and ADD 2022 challenges.  \blackf{In addition}, although we provide the unseen test set to estimate the performance of the detection models, unseen acoustic conditions and unknown fake types can degrade the performance of the detection models. So More better features and stronger methods \blackf{should} be proposed to make the detection models obtain better performance. Better approaches also need to be studied to generalize well to unknown fake utterances and acoustic environments, such as continual learning and representation learning etc.

\item{\textcolor[rgb]{0,0,0}{Reasonable evaluation metrics:}} In this research, \blackf{the EER and t-DCF}, which are employed in the previous challenges, are used as the evaluation metric in our work. However, we need to assess whether the EER \blackf{and t-DCF} are reasonable for the audio scene manipulation detection model or not in the future. We should consider human detection capabilities as well as the differences between humans and machines for detecting fake audios.

\item{\textcolor[rgb]{0,0,0}{Interpretable fake utterances analysis:}} The aim of the current work is to distinguish the manipulated audio from the bona fide one. In addition, interpretability of detection results is needed to provide in real applications. It is nontrivial to know why the utterance is fake and \blackf{find the source scene of the original audio.} \blackf{In addition}, it is also important to know what manipulation technologies are employed and even intention of the manipulation. It is particularly critical for audio forensics.
\end{itemize}

\section{Conclusions}

\blackf{Over the last few decades, a lot of important and impressive datasets have been designed for fake audio detection. However,
as far as we are aware of, the existing literature in the area of fake audio detection considers} fake types mainly including: impersonation, speech synthesis, voice conversion, replay and audio manipulation. The fake utterances \blackf{are} mostly generated by altering timbre, prosody, linguistic content or channel noise of the original audio, \blackf{which don not cover scene-fake audio}: the original scene of the utterance is manipulated by another scene. \blackf{We design the first dataset named SceneFake that considers a counterfeit method where a scene of the audio is forged by another scene using speech enhancement technologies. This paper presents the design policy, collection of real utterances, manipulation of fake audio and evaluation metrics of the SceneFake dataset. The fake audio is enhanced by using several speech enhancement models. We provide the label information of speech enhancement methods used in the fake utterances to researchers. We also report the baseline results of scene fake audio detection on our dataset. The results show that it is more challenging to detect unknown scene manipulated audio compared with the known one. We strongly believe that the publicly available SceneFake dataset and benchmark results will not only facilitate reproducible research but also further accelerate and foster research on fake audio detection and audio forensics. Our work is preliminary, and some limitations still exist, as mentioned in Section~\ref{sec:dis}. Future work includes hopefully addressing the aforementioned limitations, as well as further studying the relationship between speech enhancement performance and fake audio detection results. }

\section*{Acknowledgments}
\textcolor[rgb]{0,0,0}{This work is supported by the National Natural Science Foundation of China (NSFC) (No. 62322120, No. 62306316, No.U21B2010, No.62101553, No. 62206278).}


\bibliography{mybib}

\begin{thebibliography}{10}
\expandafter\ifx\csname url\endcsname\relax
  \def\url#1{\texttt{#1}}\fi
\expandafter\ifx\csname urlprefix\endcsname\relax\def\urlprefix{URL }\fi
\expandafter\ifx\csname href\endcsname\relax
  \def\href#1#2{#2} \def\path#1{#1}\fi

\bibitem{Zhao2013Aud}
H.~Zhao, H.~Malik, Audio recording location identification using acoustic environment signature, IEEE Transactions on Information Forensics and Security 8~(11) (2013) 1746--1759.

\bibitem{Ma2006Acou}
L.~Ma, B.~P. Milner, D.~Smith, Acoustic environment classification, ACM Transactions on Speech and Language Processing 3~(2) (2006) 1--22.

\bibitem{pandey2019n}
A.~Pandey, D.~Wang, A new framework for cnn-based speech enhancement in the time domain, IEEE/ACM Transactions on Audio, Speech, and Language Processing 27~(7) (2019) 1179--1188.

\bibitem{Pegg2024RTFSNetRT}
S.~Pegg, K.~Li, X.~Hu, Rtfs-net: Recurrent time-frequency modelling for efficient audio-visual speech separation, in: The Twelfth International Conference on Learning Representations (ICLR), 2024.

\bibitem{Stowell2015Detection}
D.~Stowell, D.~Giannoulis, E.~Benetos, M.~Lagrange, M.~Plumbley, Detection and classification of acoustic scenes and events, IEEE Transactions on Multimedia 17~(10) (2015) 1733--1746.

\bibitem{Owens2018AV}
A.~Owens, A.~A. Efros, Audio-visual scene analysis with self-supervised multisensory features, in: European Conference on Computer Vision (ECCV), 2018.

\bibitem{Gan2019Self}
C.~Gan, H.~Zhao, P.~Chen, D.~G. Cox, A.~Torralba, Self-supervised moving vehicle tracking with stereo sound, 2019, pp. 7052--7061.

\bibitem{Zhao2018TheSO}
H.~Zhao, C.~Gan, A.~Rouditchenko, C.~Vondrick, J.~H. McDermott, A.~Torralba, The sound of pixels, in: European Conference on Computer Vision (ECCV), 2018.

\bibitem{Malik2013Acoustic}
H.~Malik, Acoustic environment identification and its applications to audio forensics, IEEE Transactions on Information Forensics \& Security 8~(11) (2013) 1827--1837.

\bibitem{Zakariah2017D}
M.~Zakariah, M.~K. Khan, H.~Malik, Digital multimedia audio forensics: past, present and future, Multimedia Tools and Applications 77 (2016) 1009--1040.

\bibitem{Wu2015Spoofing}
Z.~Wu, N.~Evans, T.~Kinnunen, J.~Yamagishi, F.~Alegre, H.~Li, Spoofing and countermeasures for speaker verification: A survey, Speech Communication 66 (2015) 130--153.

\bibitem{2021ASVspoof}
J.~Yamagishi, X.~Wang, M.~Todisco, M.~Sahidullah, J.~Patino, A.~Nautsch, X.~Liu, K.-A. Lee, T.~H. Kinnunen, N.~W.~D. Evans, H.~Delgado, Asvspoof 2021: accelerating progress in spoofed and deepfake speech detection, in: he ASVspoof 2021 Workshop, 2022.

\bibitem{Yi2022ADD}
J.~Yi, R.~Fu, J.~Tao, S.~Nie, H.~Ma, C.~Wang, T.~Wang, Z.~Tian, Y.~Bai, C.~Fan, S.~Liang, S.~Wang, S.~Zhang, X.~Yan, L.~Xu, Z.~Wen, H.~Li, Add 2022: the first audio deep synthesis detection challenge, 2022, pp. 9216--9220.

\bibitem{Yi2023ADD}
J.~Yi, J.~Tao, R.~Fu, X.~Yan, C.~Wang, T.~Wang, C.~Y. Zhang, X.~Zhang, Y.~Zhao, Y.~Ren, L.~Xu, J.~Zhou, H.~Gu, Z.~Wen, S.~Liang, Z.~Lian, S.~Nie, H.~Li, Add 2023: the second audio deepfake detection challenge, in: DADA@IJCAI, 2023.

\bibitem{Kinnunen2017ASVspoof}
T.~Kinnunen, M.~Sahidullah, H.~Delgado, N.~E. M.~Todisco, et~al., The asvspoof 2017 challenge: Assessing the limits of replay spoofing attack detection, in: Annual Conference of the International Speech Communication Association (Interspeech), 2017.

\bibitem{Yi2021Half}
J.~Yi, Y.~Bai, J.~Tao, Z.~Tian, C.~Wang, T.~Wang, R.~Fu, Half-truth: A partially fake audio detection dataset, in: Annual Conference of the International Speech Communication Association (Interspeech), 2021, pp. 1654--1658.

\bibitem{Lau2004Impdataset}
Y.~W. Lau, M.~Wagner, D.~Tran, Vulnerability of speaker verification to voice mimicking, 2004, pp. 145--148.

\bibitem{Hautamaki2013Impdataset}
R.~G. Hautamaki, T.~Kinnunen, V.~Hautamaki, T.~Leino, A.~M. Laukkanen, I-vectors meet imitators: on vulnerability of speaker verification systems against voice mimicry, in: Annual Conference of the International Speech Communication Association (Interspeech), 2013.

\bibitem{Wu2015SAS}
Z.~Wu, A.~Khodabakhsh, C.~Demiroğlu, J.~Yamagishi, D.~Saito, T.~Toda, S.~King, Sas: A speaker verification spoofing database containing diverse attacks, 2015, pp. 4440--4444.

\bibitem{Wu2015ASVspoof}
Z.~Wu, T.~Kinnunen, N.~Evans, J.~Yamagishi, C.~Hanilc¸i, et~al., Asvspoof 2015: the first automatic speaker verification spoofing and countermeasures challenge, in: Annual Conference of the International Speech Communication Association (Interspeech), 2015.

\bibitem{ASVspoof2019}
A.~Nautsch, X.~Wang, N.~W.~D. Evans, T.~H. Kinnunen, V.~Vestman, M.~Todisco, H.~Delgado, M.~Sahidullah, J.~Yamagishi, K.-A. Lee, Asvspoof 2019: Spoofing countermeasures for the detection of synthesized, converted and replayed speech, IEEE Transactions on Biometrics, Behavior, and Identity Science 3 (2021) 252--265.

\bibitem{Galina2019Phone}
G.~Lavrentyeva, S.~Novoselov, M.~Volkova, Y.~N. Matveev, M.~D. Marsico, Phonespoof: A new dataset for spoofing attack detection in telephone channel, 2019, pp. 2572--2576.

\bibitem{Reimao2020For}
R.~Reimao, V.~Tzerpos, For: A dataset for synthetic speech detection, in: 2019 International Conference on Speech Technology and Human-Computer Dialogue (SpeD), 2019, pp. 1--10.
\newblock \href {https://doi.org/10.1109/SPED.2019.8906599} {\path{doi:10.1109/SPED.2019.8906599}}.

\bibitem{Frank2021WaveFake}
J.~Frank, L.~Schnherr, Wavefake: A data set to facilitate audio deepfake detection, in: NeurIPS (Benchmark and Dataset Track), 2021.

\bibitem{zang2024singfake}
Y.~Zang, Y.~Zhang, M.~Heydari, Z.~Duan, Singfake: Singing voice deepfake detection, in: Proc. IEEE International Conference on Acoustics, Speech and Signal Processing (ICASSP), IEEE, 2024, pp. 1--5.

\bibitem{Malik2013Aco}
H.~Malik, Acoustic environment identification and its applications to audio forensics, IEEE Transactions on Information Forensics \& Security 8~(11) (2013) 1827--1837.

\bibitem{Heittola2020}
T.~Heittola, A.~Mesaros, T.~Virtanen, Acoustic scene classification in dcase 2020 challenge: generalization across devices and low complexity solutions, in: Proceedings of the Detection and Classification of Acoustic Scenes and Events 2020 Workshop (DCASE 2020), 2020, pp. 56--60.

\bibitem{heittola2022TAU}
T.~Heittola, A.~Mesaros, T.~Virtanen, {TAU Urban Acoustic Scenes 2022 Mobile, Development dataset} (Mar. 2022).
\newblock \href {https://doi.org/10.5281/zenodo.6337421} {\path{doi:10.5281/zenodo.6337421}}.

\bibitem{Tan2017Speech}
Kolbaek, Morten, Tan, Zheng-Hua, Jensen, Jesper, Speech intelligibility potential of general and specialized deep neural network based speech enhancement systems, IEEE/ACM Transactions on Audio Speech \& Language Processing (2017).

\bibitem{reddy2020is}
C.~K.~A. Reddy, V.~Gopal, R.~Cutler, E.~Beyrami, R.~Cheng, H.~Dubey, S.~Matusevych, R.~Aichner, A.~Aazami, S.~Braun, P.~Rana, S.~Srinivasan, J.~Gehrke, The interspeech 2020 deep noise suppression challenge: Datasets, subjective testing framework, and challenge results, in: Annual Conference of the International Speech Communication Association (Interspeech), 2020.

\bibitem{Danieal2018Unet}
D.~Stoller, S.~Ewert, S.~Dixon, Wave-u-net: A multi-scale neural network for end-to-end audio source separation, 2018.

\bibitem{Tan2019GCRN}
K.~Tan, D.~L. Wang, Learning complex spectral mapping with gated convolutional recurrent networks for monaural speech enhancement, IEEE/ACM Transactions on Audio, Speech, and Language Processing 28~(1) (2019) 380--390.

\bibitem{Loizou2007Speech}
P.~C. Loizou, Speech enhancement: Theory and practice, CRC Press, Inc. (2007).

\bibitem{Wu2020LCNN}
Z.~Wu, R.~K. Das1, J.~Yang, H.~Li, Light convolutional neural network with feature genuinization for detection of synthetic speech attacks, in: Annual Conference of the International Speech Communication Association (Interspeech), 2020.

\bibitem{Jung2020Rawnet2}
J.~W. Jung, S.~B. Kim, H.~J. Shim, J.~H. Kim, H.~J. Yu, Improved rawnet with filter-wise rescaling for text-independent speaker verification using raw waveforms, in: Annual Conference of the International Speech Communication Association (Interspeech), 2020.

\bibitem{Two2021Li}
A.~Li, W.~Liu, C.~Zheng, C.~Fan, X.~Li, Two heads are better than one: A two-stage complex spectral mapping approach for monaural speech enhancement, IEEE/ACM Transactions on Audio, Speech, and Language Processing 29 (2021) 1829--1843.

\bibitem{Rix2002Pesq}
A.~W. Rix, M.~P. Hollier, A.~P. Hekstra, J.~G. Beerends, Perceptual evaluation of speech quality (pesq) the new itu standard for end-to-end speech quality assessment: Part i: Time-delay compensation, Journal of the Audio Engineering Society 50~(10) (2002) 755--764.

\bibitem{Taal2010ASO}
C.~H. Taal, R.~C. Hendriks, R.~Heusdens, J.~R. Jensen, A short-time objective intelligibility measure for time-frequency weighted noisy speech, in: IEEE International Conference on Acoustics, Speech and Signal Processing (ICASSP), 2010, pp. 4214--4217.

\end{thebibliography}

\end{document}